\documentclass[twocolumn,aps,prb,graphicx,showpacs]{revtex4}%
\usepackage{amsfonts}
\usepackage{amsmath}%
\usepackage{amssymb}%
\usepackage{graphicx}

\begin{document}

\title{Dynamical Mean Field Theory of Double Perovskite Ferrimagnets}
\author{K. Phillips$^1$, A. Chattopadhyay$^2$, A. J. Millis$^3$}
\affiliation{
$^1$Department of Physics and Astronomy, Rutgers University\\
136 Frelinghuysen Rd, Picataway, NJ 08854\\
$^2$IBM Almaden Research Center\\
650 Harry Road, San Jose, CA\\
$^3$Department of Physics, Columbia University\\
538 West 120th Street, NY, NY, 10027\\}
\date{\today}

\begin{abstract}
The dynamical mean field method is used to analyze
the magnetic transition temperature and optical conductivity
of a model for the ferrimagnetic double perovskites
such as $Sr_2FeMoO_6$. The calculated transition temperatures
and optical conductivities are found to depend sensitively
on the band structure. For parameters consistent with
local spin density approximation band calculations, the computed
transition temperatures are lower than observed, and in 
particular decrease dramatically as band filling is increased,
in contradiction to experiment. Band parameters which would 
increase the transition temperature are identified.
\end{abstract}


\maketitle

\section{Introduction}

The calculation of non-zero-temperature
and dynamical properties,  such as magnetic 
transition temperatures and conductivities, is a long-standing and
difficult problem in materials theory, but one for which the recent theoretical
development of the 'dynamical mean field' method \cite{Georges96} offers a
promising avenue for progress. This method, which allows an exact (in
principle) treatment of local correlations has been recently used in
combination with 'ab initio' band calculations to estimate the transition
temperatures of $Fe$ and $Ni$ \cite{Fe} , and in combination with a tight
binding parametrization of band theory to elucidate the physics of the
'colossal' magnetoresistance manganites 
\cite{Furukawa95,Millis96b,Quijada98,Chattopadhyay00}
and models of magnetic semiconductors \cite{Chattopadhyay01b,Hwang02}.

In
this paper we apply the method to study ferrimagnetic members
of the 'double perovskite' family of
materials. These are compounds of chemical formula $ABB^{\prime}O_{6}$, with
$A$ an alkaline earth such as $Sr$, $Ca$ or $Ba$, and $B,B^{\prime}$ two
different transition metal ions. In the ferrimagnets of present
interest the $B$ site is occupied by  $Fe$ and $B'$ site
by a member of the $4d$ or $5d$ transition metal series such
as $Re$ or  $Mo$. The double perovskite family of  materials
have long been known \cite{Sleight68} but the ferrimagetic
members listed above have become the subject of recent
interest \cite{Kobayashi98,Terakura99,Sarma00,Gopalakrishnan00,Ray01}
because they seem to be to be metallic (except perhaps in the
$Ca$ case \cite{Gopalakrishnan00} ferrimagnets with magnetic transition
temperatures greater than room temperature and apparently highly
spin-polarized conduction bands, raising the possibility of interesting device
applications as 'spin valves' \cite{spinvalve}, elements in magnetic
information storage systems \cite{media} and as sources of spin polarized
electrons for spintronic applications \cite{spintronic}. 

The materials are
also of fundamental interest, because their physics and materials 
science are
far from understood. For example, apparently minor changes in chemical
composition or processing conditions can change the electrical behavior from
metallic to insulating, or the magnetic transition temperature from $400K$ to
$\ 200K$. More generally, the materials provide examples of novel behaviors
associated with partial filling of transition metal $4d/5d$ shells, which
have been little studied in comparison to the $3d$ transition metal series
\cite{TokuraRMP}. Further, the specific materials we wish to study
are believed to be half metallic ferrimagnets, but the obvious strong
interaction which could give rise to half
metallicity (a Hunds coupling on the $Fe$) exists on only
one of the two sites, unlike the situation in the somewhat analogous
CMR materials. Thus behavior arising from the interplay of magnetic
order and carrier motion may be different.
Finally, the origin of the magnetism is not settled.
A natural guess is that it is due to
the strong Hunds coupling on the $Fe$ site, but other interactions
have been proposed to be important \cite{Sarma00}

In this paper we present a model for the low (less than, say $3eV$)
lying  electronic states, which are important for transport and magnetism.
Our model consists of a tight binding description of the bands,
which we derive from  previously published first principles calculations
\cite{Terakura99,Sarma00} and general arguments, 
and a local interaction (namely a
Hunds coupling on the $Fe$ site). The key physical assumption
made in our model is that the magnetism is driven by the
strong Hunds coupling on the $Fe$ site. Important technical
issues include the two dimensional nature of the underlying
band structure and the multiorbital nature of the material.
We
solve the model in the dynamical mean field approximation, and from our
solution determine the magnetic transition temperature and optical conductivity,
and attempt to determine the general materials aspects which control $T_c$.
This paper supercedes a previous
paper \cite{Chattopadhyay01a}, in which the model Hamiltonian used did not
provide an adequate approximation to the underlying band structure. 

\section{Material and Model}

\subsection{Material}

Double perovskite systems form in the $ABB^{\prime}O_{6}$ crystal structure
which generalizes the $ABO_{3}$ perovskite structure familiar from
ferroelectrics, high temperature superconductors and the 'colossal'
magnetoresistance rare earth manganites by  having two different $B$ site
ions.  In the double perovskite materials of  interest here,\ $A$ is an
alkaline earth such as $Sr$, $Ca$ or $Ba$ and the $B,B^{\prime}$ sites form a
rocksalt structure, i.e. a simple cubic lattice with a doubled unit cell and
one sublattice occupied by $Fe$ and the other by a transition metal from the
$4d$ or $5d$ series such as $Mo$ or $Re$. The crystal fields
and atomic energetics are such that the formal valences correspond to $Fe$
with a half filled, maximally polarized $d$-shell while the
$Mo/\operatorname{Re}$ has one or two d electrons distributed over the
$t_{2g}$ levels \cite{Sleight68,Terakura99,Sarma00}. 
We will focus on electronic
states arising from the transition metal d-levels.

\subsection{Hamiltonian}

\subsubsection{Overview}

The Hamiltonian describing the low lying, electronically active degrees of
freedom may be written as the sum of a 'hopping' part arising from the band
structure and an interaction part:
\begin{equation}
H=H_{band}+H_{int}\label{H}%
\end{equation}
The relevant portions of the calculated \cite{Terakura99,Sarma00} 
band structure
involve three bands (degenerate in the ideal double perovskite  structure)
arising from the three transition metal $t_{2g}$ levels $d_{xy,yz,xz}$ To a
high degree of accuracy these three bands do not hybridize with each other and
the physics is therefore described by a three-fold degenerate tight binding
model. The planar character of the $t_{2g}$ levels implies that the tight
binding model has an interesting two dimensionality, which may be summarized
as follows. The $d_{xy}$ orbital on a $Fe$ site hybridizes via a matrix
element $t_{1}$ with the $d_{xy}$ levels on the four nearest neighbor
($Mo/Re$), sites {\it in the same plane} and via a much
smaller matrix element $t_{3}$ to the four nearest $Fe$ ions also in the same
plane. The hopping in the third direction is negligible, because of the planar
character and $xy$ orbital symmetry of the $d_{xy}$ wave function. The
$d_{xy}$ orbital on a $Mo/\operatorname{Re}$ site hybridizes with the four
in-plane near neighbor $Fe$ sites via the same hopping matrix element $t_{1}$
and with the four in-plane second nearest neighbor ($Mo/\operatorname{Re})$
sites, via another matrix element $t_{2},$ which is not particularly small,
because of the more spatially extended character of the $d$-electrons in
$4d/5d$ orbitals.   Further neighbor hoppings are also found to be important
in other $t_{2g}$-based $4d$ systems such as $Sr_{2}RuO_{4}$
\cite{Ruthenateband}.  

It is natural to assume that the magnetic character of the material
derives from the strongly magentic nature of the $Fe$
ion and we therefore  assume that the 
dominant interaction arises from the strong atomic Hunds
coupling of the $Fe$.

\subsubsection{Hopping Hamiltonian}

To write the Hamiltonian explicitly we focus the cubic lattice of
$B,B^{\prime}$ sites in the underlying single perovskite structure, labelling
these sites by $i$ and the operator creating an electron into orbital
$a(=xy,yz,xz)$ and spin $\sigma$ by $c_{a,i,\sigma}^{+}$. \ Although we refer
to this orbital as a $^{\prime}d$-orbital' it  in fact represents a hybrid,
composed mainly of transition metal $d$ and oxygen $p$ orbitals, of the
correct local symmetry. We introduce a nearest neighbor ($Fe\leftrightarrow
Mo/\operatorname{Re}$) hopping $t_{1}$ and two second neighbor (same
sublattice) hoppings $t_{2}$ and $t_{3}$ representing $Mo-Mo$ or $Fe-Fe$
hoppings respectively. As noted above we expect that $t_{2}$ corresponding to
to $Mo-Mo$ hopping is relatively large, because of the larger spatial extent
of the $4d/5d$ orbitals while  $t_{3}$ is essentially negligible. To obtain
the conductivity we couple in the electric field by using a vector potential
and the Peierls phase ansatz. This approximation has been shown to be accurate
in other transition metal oxide contexts \cite{Ahn00,Millis01a}.Thus the
hopping portion of the Hamiltonian is the sum of three identical tight binding
models, one for each orbital. The Hamiltonians take the general form (note
that the first sum runs over all lattice sites, the second over the
$B^{\prime}$ (non-Fe) and the third over the B (Fe) sites, while $\delta_{a}$
labels the in-plane direction relevant to orbital $a$ and we have set the
electric charge $e$ and the speed of light $c$ equal to unity)
\begin{align}
H_{band} =& -\sum_{a,i,\delta_{a},\sigma}\left(  t_{1,a}e^{i\mathbf{A\cdot
\delta}_{a}}c_{a,i,\sigma}^{+}c_{a,i+\delta_{a},\sigma}+H.c.\right)
\nonumber\\
&  -\sum_{a,i\in B^{\prime},\delta_{a}^{\prime}}\left(  t_{2,\alpha
}e^{i\mathbf{A\cdot\delta}_{a}^{\prime}}c_{i,a,\sigma}^{+}c_{i+\delta
_{a}^{\prime},a,\sigma}+H.c\right)  \label{Hband}\\
&  -\sum_{a,i\in B,\delta_{a}^{\prime}}\left(  t_{3,\alpha}e^{i\mathbf{A\cdot
\delta}_{a}^{\prime}}c_{i,a,\sigma}^{+}c_{i+\delta_{a}^{\prime},a,\sigma
}+H.c\right)  \nonumber
\end{align}

$H_{band}$ implies an interesting band structure, which is most plainly
revealed by writing $H_{band}$ in momentum space in a matrix notation where
the upper left entry corresponds to $Fe$ and the lower right to $Mo$, thus if
$A=0$ we have, for the $xy$ orbitals
\begin{align}
H_{band,xy} & [A=0]= \label {Hbandxy}\\
&\left(
\begin{array}
[c]{cc}%
 0& -2t_{1}\left(  \cos p_{x}+\cos p_{y}\right)  \\
-2t_{1}\left(  \cos p_{x}+\cos p_{y}\right)   & -4t_{2}\cos p_{x}\cos p_{y}%
\end{array} 
\right) \nonumber
\end{align}
where we have set the $Fe-Mo$ distance to unity and the momenta are restricted
to the reduced Brillouin zone $\left\vert p_{x}\right\vert +\left\vert
p_{y}\right\vert <\pi.$

\subsubsection{Interaction}

The most important interaction effect constrains the occupancy of the $B$
($Fe)$ site.  The formal valence of $Fe$ is $d^{5}$ and the strong Hunds
coupling characteristic of $Fe$ (and found in the local spin density
approximation to band theory) means that
in the $d^5$ configuration all of the $Fe$ $d$-electrons are
aligned, leading to a filled, completely spin-polarized $d$-shell. Two charge
fluctuation processes are possible: $Fed^{5}\leftrightarrow Fed^{4}$ or
$Fed^{5}\leftrightarrow Fed^{6}$ The strongly stable nature of the filled
d-shell implies that the $d^{5}-d^{6}$ process is dominant. To express this
physics we introduce a strong Hunds coupling on the $B$ ($Fe$) site,
expressing the fact that in the ground state the $Fe$ is in the $d^{5}$
maximal spin configuration, and an energy splitting parameter $\Delta$
expressing the differing electronegativities of the $B$ and $B^{\prime}$
sites. Thus we write
\begin{align}
H_{int}=&-J\sum_{a,i\in B\alpha\beta}\overrightarrow{S}_{i}\cdot c_{a,i,\alpha
}^{+}\overrightarrow{\sigma}_{\alpha\beta}c_{a,i,\beta}+ \nonumber \\
&\sum_{a,i\in
B^{\prime},\sigma}\Delta^{a}c_{a,i,\sigma}^{+}c_{a,i,\sigma}\label{hint}%
\end{align}
For the calculations presented in this paper we will specialize to cubic
symmetry, so the $\Delta$ are the same for all three orbitals, but this
restriction may easily be lifted. The energy scale relevant for the
$Fed^{5}\leftrightarrow Fed^{6}$ valence fluctuation is $J-\Delta.$
Examination of published band structures \cite{Terakura99,Sarma00} indicates that
$J-\Delta \sim 1eV$ while the $d^{5}\leftrightarrow d^{4}$ process has a
much larger energy of $\left\vert J+\Delta\right\vert \geq5eV.$ 
In a fully spin polarized ground state, the interaction 
terms simply become level shifts, $\Delta_{maj}=J+\Delta$ and
$\Delta_{min}=-J+\Delta$ for the majority and minority spin
bands. Transitions onto the majority-spin $Fe$ orbital
involve very large energies, so to simplify
the calculations at 
$T>T_c$ we will take the limit $J+\Delta\rightarrow\infty$ with
$J-\Delta$ fixed. We henceforth refer to the quantity $J-\Delta$ as $\Delta$.
Because the local spin density approximation may not be accurate for strongly
interacting systems such as the double perovskites, we will consider a range
of $\Delta$ here.

Other authors \cite{Ray01} have argued that an additional Hunds-type
coupling on the $B^{\prime}$ site is important. Technical limitations prevent
us from treating such an interaction accurately, so we do not include it here.

\subsubsection{Discussion: T=0 Band structure, ferromagnetic case}

Eq \ref{Hbandxy} may be thought of as describing two bands of electrons: one
on the $Fe$ sites, with 'intrinsic' bandwidth set by $t_{3}$ 
and one on the non-$Fe$
sites, with 'intrinsic' 
bandwidth set by $t_{2}$. The two bands hybridize via the overlap
$-2t_{1}\left(  \cos p_{x}+\cos p_{y}\right)  $.  We see immediately that  the
hybridization vanishes along the line $\cos p_{x}+\cos p_{y}=0$ which also
contains the van Hove points $0,\pi$ and $\pi,0$ at which the density of
states of the two individual bands diverges.  Near these
points a complicated structure
including divergences in the density of states is expected. 

\begin{figure}
\includegraphics[width=3.0in]{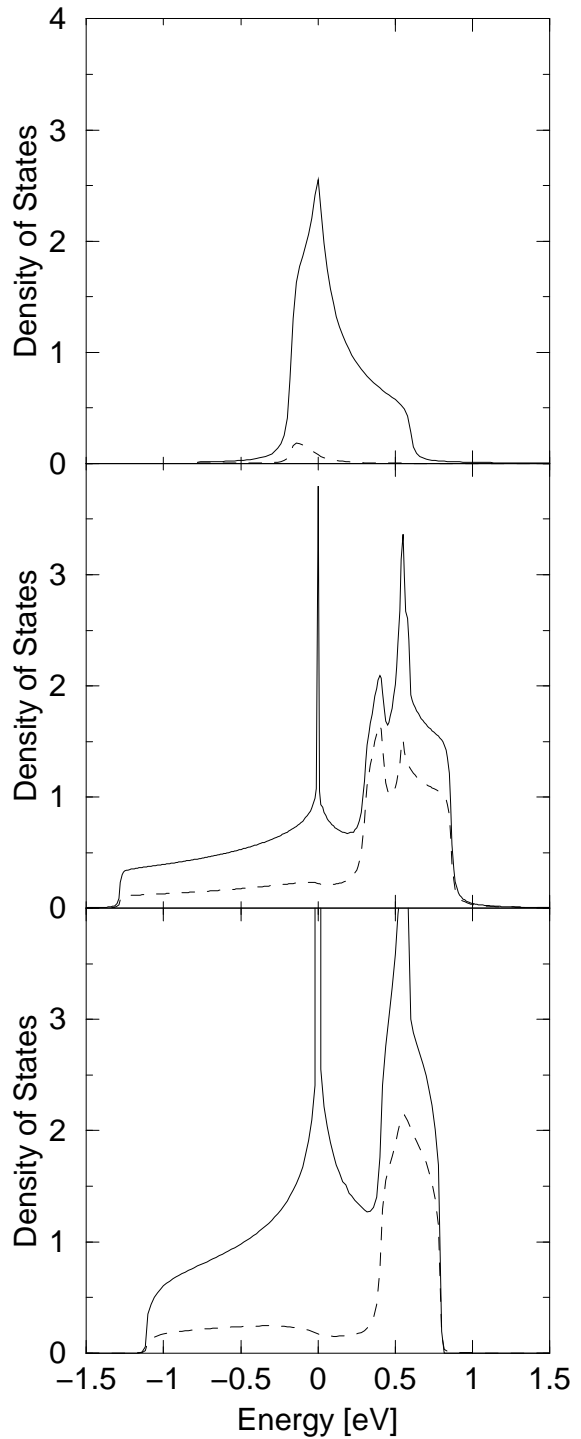}

\caption{ Total (solid line) and $Fe$-projected (dashed line) densities of
states computed  using 
tight binding parameters $t_1=0.25$ eV, $t_2=0.15eV$, $t_3=0.03eV$.
Top panel: $T=0$ majority spin density of states 
using $\Delta_{maj}=-2.5$eV
Middle panel: $T=0$ minority spin density of
states computed from Eq. \ref{Hbandxy} with $\Delta_{min}=0.3$eV.
Lowest panel: total (both spins) density of states at $T>T_c$ computed 
as described in section III.}
\end{figure}

The full density of states and the projection of this
density of states onto the $Fe$ orbitals are shown in 
the upper panels of Fig. 1 for parameters 
$t_1=0.25$  eV, $t_2=0.15eV$, $t_3=0.03eV$
(note that most of the majority spin
$Fe$ density of states is at a low energy outside the range
of this plot).
Comparison of this density of states
to the published band theory results \cite{Terakura99,Sarma00}
shows that these parameters reproduce the band density of
states accurately. The main difference is that if the
$J$ and $\Delta$ are adjusted to correctly reproduce the minority
spin band, then the upper (non-$Fe$-portion of the majority spin band
is positioned about $0.5eV$ too low in energy.
The extra shift in the majority spin $Mo$ orbitals must be attributed
to a Hunds coupling on the non-Fe site, not included here.

The two  features seen in our calculated density of states   near $0.5eV$
arise from \ states in the vicinity of the van Hove points $(0,\pi)$ and
$(\pi,0)$ where the hybridization vanishes and the $B$ and $B^{\prime}$ sites
have energy $\Delta+4t_{3}\simeq0.4eV$ and $\ 4t_{2}\approx0.6eV$
respectively$,$ whereas the peak at $\omega=0$ arises from the van Hove point
$(\pi/2,\pi/2)$ of the $B^{\prime}$ (non-Fe) band, where
as noted above the hybridization to the $Fe$ vanishes.

Formal valence arguments indicate that the material contains
one or two d-electrons beyond the filled shell $Fe-d^5$ $Re/Mo-d^0$
configuration. Fig 1 shows that for the band theory parameters,
these carriers go into states with only a small
admixture of $Fe$. The physics
behind this result is that for this sign of $t_2$
the strongly hybridized states near $p_{x}=p_{y}=0$ are at the bottom of the
band described by the $t_{2}-$only term in $H_{band}$, 
and are pushed further away from the $Fe$ states by
usual level repulsion, leading to a mainly non-Fe character of
the lowest states.
For $n-1$ ($Re$) only the minority
spin band is occupied (the majority spins occupy 
low-lying $Fe$ states off of the plotted scale).
However, for $n=2$ within this approximation, chemical
potential is $\mu_2 \approx 0$ and the majority spin
band is somewhat occupied, so the material is not a 'half-metal'
in this approximation.
These features will be seen to be of importance for the
calculated transition temperature and optical conductivity.

\begin{figure}
\includegraphics[width=3.0in]{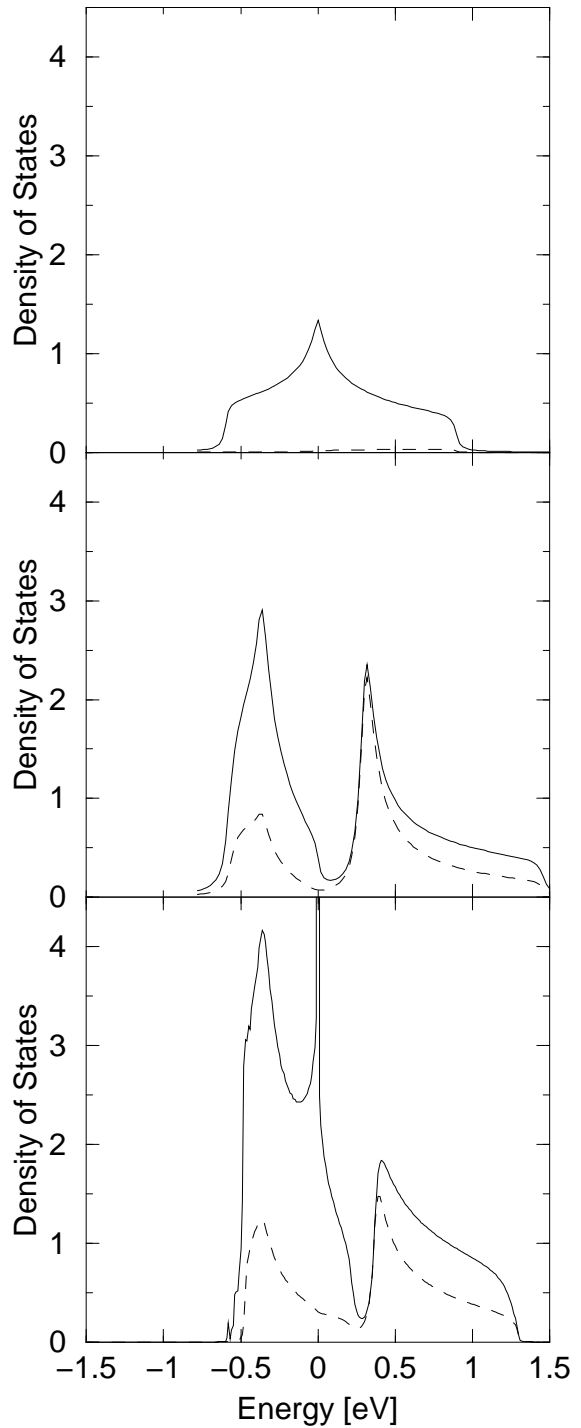}
\caption{Total (solid line) and $Fe$-projected (dashed line) densities of
states computed  using alternative
tight binding parameters $t_1=0.25$ eV, $t_2=-0.15eV$, $t_3=0.03eV$.
Top panel: $T=0$ 
majority spin density of states computed from Eq. \ref{Hbandxy}
using $\Delta_{maj}=-2.5$eV.
Middle panel: $T=0$ minority spin density of
states computed from Eq. \ref{Hbandxy} with $\Delta_{min}=0.3$eV.
Lowest panel: total (both spins) density of states at $T>T_c$ computed 
as described in section III.}
\end{figure}

It is interesting to consider a contrasting set of parameters,
for which the level repulsion argument works in the opposite
manner. If $t_2$ has an unphysical (negative) sign and $\Delta$ is
near $0$ then the low lying states of the $t_2$-band do
not mix with the $Fe$ states, which are pushed downwards
by hybridization with the higher-lying $Re/Mo$ levels,
leading to low-lying states of mainly $Fe$ character,
as shown in Fig 2.

\subsection{Conductivity}

The current operator $\widehat{J}=\delta H/\delta A$ \cite{Millis01a}. For
electric field in the $x$ direction the $xy$ and $xz$ orbitals contribute,
thus $J_{x}=J_{x,xy}+J_{x,xz}$ with $\delta_{x}$ the lattice vector in the $x$
direction and $\delta^{\prime}$ \ labelling the four 'second neighbor' lattice
vectors ($\delta^{\prime}=\pm\left(  \mathbf{\delta}_{x}\pm\mathbf{\delta}%
_{y}\right)  )$ so by expanding Eq \ref{Hband} in powers of $A$ we obtain
\begin{align}
J_{x,xy}(A) &  = \nonumber \\
&-\sum_{i,\pm\delta_{x},\sigma}\left(  i\delta_{x}%
t_{1,a}e^{i\mathbf{A\cdot\delta}}c_{xy,i,\sigma}^{+}c_{xy,i+\delta_{x},\sigma
}-H.c.\right)  -\label{j}\\
&  \sum_{i\in B^{\prime},\delta_{a}^{\prime}}\left(  it_{2,\alpha}\left(
\delta^{\prime}\cdot\widehat{\mathbf{x}}\right)  e^{i\mathbf{A\cdot\delta}%
_{a}^{\prime}}c_{i,a,\sigma}^{+}c_{i+\delta_{a}^{\prime},a,\sigma}-H.c\right)
\nonumber
\end{align}
The expectation value of the term in $J$ proportional to $A$ gives the total
oscillator strength, $S(\infty)=\frac{2}{\pi}\int_{0}^{\infty}d\omega
\sigma_{1}(\omega)$ in the conduction band contribution to the optical
conductivity (see \cite{Maldague77,Quijada98,Millis01a} for details). 
Restoring
units we have ($a$ is the $Fe-Mo/\operatorname{Re}$ distance, the sum rule is
conventionally defined in terms of the conductivity per unit volume and the
factor of two comes from the $xy$ and $xz$ orbitals, whose contributions to
the expectation values are identical in cubic symmetry)%
\begin{align}
S(\infty)=&\frac{2e^{2}}{a}\left\langle \sum_{i,\pm\delta_{x},\sigma}\left(
t_{1,a}c_{xy,i,\sigma}^{+}c_{xy,i+\delta_{x},\sigma}+H.c.\right)
\right\rangle \\ \label{Sinf}
&+\left\langle \sum_{i\in B^{\prime},\delta_{a}^{\prime}}\left(
t_{2,\alpha}c_{i,a,\sigma}^{+}c_{i+\delta_{a}^{\prime},a,\sigma}+H.c\right)
\right\rangle \nonumber
\end{align}
The conductivity is
\begin{equation}
\sigma(\Omega)=\frac{S(\infty)-2\chi_{jj}(\Omega)}{i\Omega}%
\end{equation}
with $\chi_{jj}$ the usual Kubo formula current-current correlation function
evaluated using $J_{x}(A=0)$ (Eq \ref{j}) and again the factor of two
represents the orbital degeneracy.

\section{Method of Evaluation}

\subsection{Overview}

To evaluate the properties of $H$ we use the dynamical mean field method
\cite{Georges96,Chattopadhyay00} This method is extensively described and
justified elsewhere, and is relevant here because the principal interactions
are local. In brief the central approximation is that the electron self
energy, $\Sigma$, is momentum independent. \ In this circumstance the physics
may be derived from a local theory which may be viewed as a quantum impurity
model combined with a self consistency condition. The application to the
double perovskite systems requires some discussion. In these systems the unit
cell contains two sites, each site contains three orbitals and there are two
choices of spin, so the local theory has twelve degrees of freedom. However,
the problem may be simplified. First, the three orbitals ($d_{xy}$ etc) are
coupled only via the interaction. Second, the interaction exists only on the
$Fe$ site, so that we may formally integrate out the electrons on the non-$Fe$
($B^{\prime}$) site, defining a single-orbital model with  the inverse $Fe$
($(B)$-site Green function for \ e.g. the $xy$ orbitals viz

\begin{align}
G_{BB}^{xy,band}(p,\omega)^{-1}=\omega
-\frac{4t_{1}%
^{2}\left(  \cos(p_{x})+\cos p_{y}\right)  ^{2}}{\omega+\Delta^{xy}%
-4t_{2}(\cos(p_{x})\cos(p_{y}))}
\label{g22band}
\end{align}

We measure momenta in units of $\pi/a$ where $a\approx4\mathring{A}$ is the
distance from a $B$ to a nearest neighbor $B^{\prime}$ site. The two
dimensional Brillouin zone is defined by $\left\vert p_{x}+p_{y}\right\vert
<\pi$.

The physics is then described by a three-orbital local theory given by the
partition function $Z_{loc}=\int\mathcal{D}c^{+}c\exp[S_{loc}]$ with an action
$S_{loc}$ which we write in the Matsubara frequency representation as
\begin{equation}
S_{loc}=T\sum_{\omega}Tr[c_{a\alpha}^{+}(\omega)\left(  \mathbf{a}%
_{\alpha\beta}^{ab}(\omega)-J\mathbf{S\cdot\sigma}_{\alpha\beta}\right)
c_{a\beta}(\omega)] \label{simp}%
\end{equation}
involving fields $c_{a\alpha}$ and specified by a tensor mean field function
$\mathbf{a}$ which has orbital ($ab$) and spin ($\alpha\beta$) indices (the
trace is over the spin and orbital indices). The mean field function is fixed
by the condition that the Green function defined from $S_{loc}$,
\begin{equation}
\mathbf{G}_{loc}(\tau)=\frac{\delta\ln Z_{loc}}{\delta\mathbf{a}(\tau
)}=\left(  \mathbf{a}-\mathbf{\Sigma}\right)  ^{-1} \label{gloc}%
\end{equation}
is equal to the local Green function defined by integrating Eq \ref{g22band}
over momenta using the self energy defined by Eq \ref{gloc} i.e.
\begin{equation}
G_{loc}^{xy}(\omega)=\int\frac{d^{2}p}{(2\pi)^{2}}G_{BB}^{xy,band}(p,\omega-\Sigma)
\label{mfeq}%
\end{equation}
and the integral is over the Brillouin zone defined above.

Substitution of Eqs \ref{simp} \ref{gloc} into Eq \ref{mfeq} yields explicit
equations which are solved numerically by iteration.

\subsection{Calculation of T$_{c}$}

We calculate the ferromagnetic transition temperature by decomposing the mean
field function $\mathbf{a}$ into non-magnetic ($a_{0}$) and magnetic $(a_{1}$)
parts
\begin{equation}
\mathbf{a}=a_{0}^{a}+a_{1}^{a}\mathbf{m\cdot\sigma}\label{a0}%
\end{equation}
and linearizing in $a_{1}$. We take the magnetization direction $\mathbf{m}$
to be parallel to $z$ and take the limit $J\rightarrow\infty$ so that after
integrating out the fermions and redefining $a\rightarrow a+J$ we obtain
($\cos(\theta)$ is the dot product between the direction of the core spin and
of the magnetization)
\begin{equation}
S_{imp}=Tr\ln\left[  a_{0}^{a}(\omega)+a_{1}^{a}(\omega)\cos(\theta)\right]
\label{simp2}%
\end{equation}
where the $Tr$ is over the frequency index and the orbital degree of freedom.

The Green function of the impurity model becomes%
\begin{equation}
G_{imp}^{a}(\omega)=\frac{1}{2}\left\langle \frac{1-\widehat{S}\cdot
\overrightarrow{\sigma}}{a_{0}-a_{1}\cos(\theta)}\right\rangle \label{gimp1}%
\end{equation}
where the expectation value is over the directions of the 'core spin' $S$.

In the paramagnetic phase $a_{1}=0$. Expanding near the magnetic transition
(assumed second order) yields
\begin{equation}
G^{imp}(\omega)=\frac{1}{2a_{0}}\left(  1-\left(  m+\frac{a_{1}}{3a_{0}%
}\right)  \sigma_{z}\right)  \label{gimp2}%
\end{equation}
with $m=<\cos\theta>$so that
\begin{equation}
\Sigma(\omega)=-a_{0}-\left(  2a_{0}m-\frac{a_{1}}{3}\right)  \sigma
_{z}\label{sigimp}%
\end{equation}

At $T>T_{c}$ $m=a_{1}=0$ and substitution of Eqs \ref{sigimp}, \ref{gimp2}
into Eq \ref{mfeq} yields
\begin{equation}
\frac{1}{2a_{0}(\omega)}=I_{1}(\omega,a_{0}(\omega))
\end{equation}
where the $n^{th}$ order integral $I_{n}$ is given by
\begin{equation}
I_{n}=\int\frac{d^{2}p}{\left(  2\pi\right)  ^{2}}\left(  G_{22}%
(\omega)\right)  ^{n}\label{Indef}%
\end{equation}
This equation is solved numerically by iteration for a sufficiently dense set
of frequency points (typically frequency spacing $0.04t_{1}$). Once a solution
for $a_{0}$ is obtained we may linearize Eq \ref{mfeq} in the magnetic part of
the self energy and local Green function, obtaining
\begin{equation}
\frac{m}{2a_{0}}-\frac{a_{1}}{6a_{0}^{2}}=I_{2}(\omega,a_{0}(\omega))\left(
2a_{0}m+\frac{a_{1}}{3}\right)
\end{equation}
where
\begin{equation}
m=<\cos\theta>=\sum_{\omega,a}\frac{a_{1}^{a}}{3a_{0}^{a}}\label{m}%
\end{equation}
Solving for $a_{1}$ and then using this to obtain an expression for $m$ yields
a self consistent equation for $T_{c}$ which in the limit of cubic symmetry
becomes%
\begin{equation}
1=\sum_{n}\frac{1-4a_{0}^{2}I_{2}}{1+2a_{0}^{2}I_{2}}=\sum_{n}\left[
-2+\frac{9}{1+2a_{0}^{2}I_{2}}\right]  \label{tceq}%
\end{equation}
It turns out that the transition temperatures are low compared to the other
scales of the model so that \ that one may recast this equation via analytical
continuation to the real axis as ($\mu$ is the chemical potential
corresponding to the desired carrier density)%
\begin{equation}
T_{c}=\int_{-\infty}^{\mu}\frac{d\omega}{\pi}\operatorname{Im}\left[  \frac
{9}{1+2a_{0}^{2}I_{2}}\right]  \label{tcfinal}%
\end{equation}

\section{Results and Discussion}

We have calculated
the magnetic transiton temperature from Eq. \ref{tcfinal}, finding that for
the parameters used to construct Fig. 1 (and which
are the ones following from band theory)
$n=1$ ($\operatorname{Re}$ case) $T_{c}\approx110K$ and that for $n=2$ ($Mo$
case) the ground state is not ferromagnetic. \ These calculated values are in
sharp disagreement with the experimental values $T_{c}\gtrapprox400K$ for both
$n=1$ and $n=2$. The relatively small values of $T_{c}$ found for $n=1$ in
this calculation\ may be understood from the density of states, which shows
that the low-lying states lie mainly on the $non-Fe$ sites, which are far
displaced in energy from the magnetic site and therefore do not hybridize
strongly with it, so the effective carrier-spin interaction is not strong.
That the  $n=2$ is non-magnetic may be understood by combining the results of
\cite{Chattopadhyay00} with the observation that the band structure is
effectively two dimensional. In the extreme weak coupling limit, the nature of
the magnetic ground state is determined by the wave vector at which the
susceptibility is maximal.  For the two dimensional band structures
considered here this maximum is not at $q=0$. Ref \cite{Chattopadhyay00}
showed that in the DMFT approximation, increasing the carrier-spin coupling
increased the range in which ferromagnetism existed, but that as band filling
is increased, a transition to an  antiferromagnetic state generically occurs,
and gets pushed to the half-filled band only for $J$ of the order of the
bandwidth. These effects are  more pronounced for the two dimensional band
structure we consider.

The relative weakness of the virtual $\operatorname{Re}/Mo\leftrightarrow Fe$
transitions is reflected in the temperature dependence of the many-body
density of states, shown for the 'band'
parameters in the lower panel of Fig. 1. Comparing these
we see  that \ disordering the $Fe$ spins leads to a slight
narrowing of the bands, but the larger ($30\%$) band narrowing effects found
in $CMR$ manganites \cite{Millis96b}  are not observed for these parameters.

\begin{figure}[ht]
\includegraphics[width=3.0in]{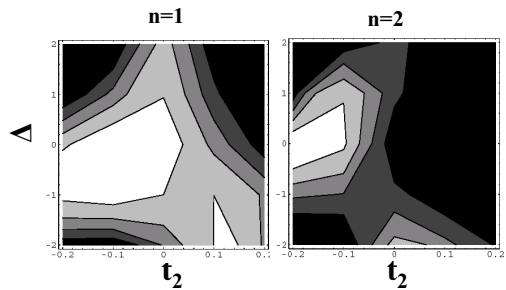}
\caption{{\it Left panel}: Contour plots of calculated transtion temperatures
for range of model parameters and conduction band
density $n=1$. 
{\it Right panel:} Contour plots of calculated
transition temperature for $n=2$.
In each figure, contours are spaced approximately
100 Kelvin apart and the white areas correspond to transition
temperatures in excess of $400K$.}
\end{figure}

To understand the behavior of the model in more detail we have evaluated the
predicted ferromagnetic transtion temperatures for  wide range of model
parameters. The results are summarized in the two panels of Fig. 3 which show
via  contour plots the values of $T_{c}$ predicted by the method. The contours
are spaced approximately $100K$ apart, and the black regions indicate the
areas in which the calculated $T_{c}$ vanishes. It is seen that in order to
obtain a reasonably high transition temperature, especially for the $n=2$ band
filling, one must choose the parameter $t_{2}$ to have the opposite sign from
the physical one. The reason for this behavior is 
reveal by Fig. 2, which shows the density of states
for parameters which maximize the $n=2$ $T_c$.
The low-lying states for this case are seen to
be of mainly $Fe$ character, because the level repulsion
argument which pushed down the non-$Fe$ states for the LSDA parameters
is not operative here.

\begin{figure}[ht]
\includegraphics[width=3.0in]{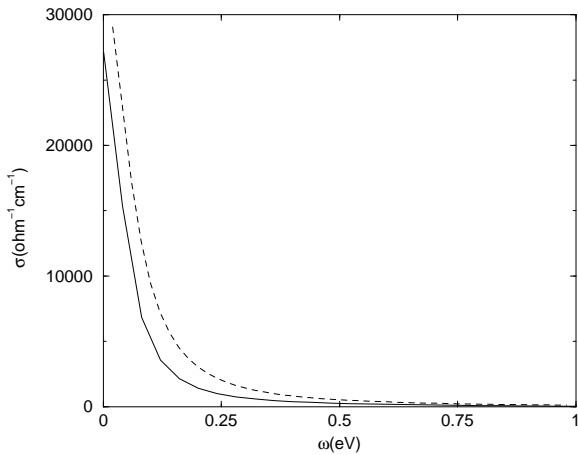}
\caption{$T=0$ (dashed) and $T>T_c$
conductivities for $n=1$ 
using best-fit band parameters $t_1=0.25eV$, $t_2=0.15eV$
$t_3=0.03eV$ and $\Delta=0.3eV$ used in Fig. 1.
The $T=0$ conductivity was computed using an artificial
broadening of $0.1eV$ applied to the $B$ site.}
\end{figure}

We have also calculated the optical
conductivity for various model parameters.
Results obtained using the LSDA parameters are
shown in Figs. 4 ($Re$-case, $n=1$)
and fig.  5 ($Mo$-case, $n=2$).
One would in principle expect
two classes of transitions: a
'Drude' peak centered at $\omega=0$ involving motion of electrons 
near the Fermi surface and an interband transition
involving moving an electron from a $Re/Mo$ to a $Fe$.
Our calculations indicate that for the tight binding
parameters corresponding to the LSDA calculation,
the interband feature is very weak, indeed not visible in the Figure
again demonstrating the weakness of the $Fe-Mo$ coupling
for these parameters. We observe that the 'Drude' part has a distinctly non-
Drude form, which arises because in our calculation the scattering processes
couple to the 'B' ($Fe$) site only; although the regions of momentum space
where the hybridization vanishes are of measure zero, they do lead to
a frequency dependence of the scattering rate which explains the peculiar
form. We also note that the main cause of the changes in
conductivity and oscillator strength between $T=0$ and $T=T_c$ is
the change in band filling, which leads to a change in 
optical matrix element. In the paramagnetic state one has three
bands, each with a two-fold spin degeneracy, corresponding to
a filling of $n/6$ (n=1,2 is the particle density), so the relevant
states are quite close to the bottom of the band where the optical matrix
element is small. In the ferrimagnetic state for $n=1$ one loses the spin 
degeneracy, so one has three bands each filled to a higher level, so with a
correspondingly higher fermi velocity and optical matrix element,
whereas for the $n=2$ case the temperature induced shift corresponds to a 
change from $1/3$ to $2/3$ filled band, with much smaller change in optical
matrix elements. The temperature dependent change in the oscillator strength is
therefore much less.
As noted above, in the ferrimagentic case for $n=2$ one has a small
filling of the majority spin band, leading to a small
additional contribution to $\sigma$, shown as the dot-dashed line in Fig. 5.

\begin{figure}
\includegraphics[width=3.0in]{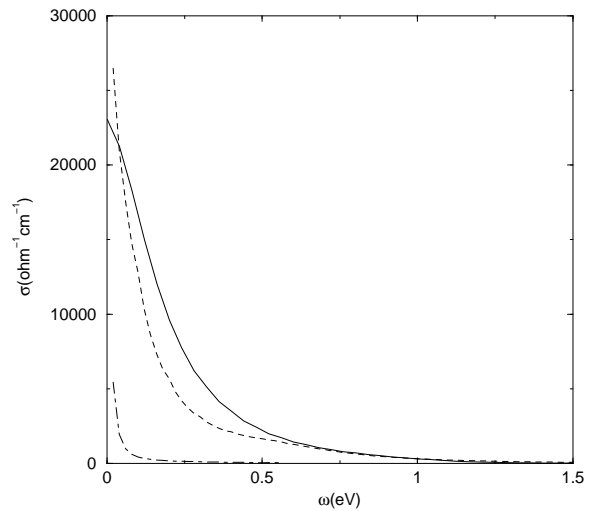}
\caption{Upper panel: $T=0$ (dashed) and $T>T_c$ (solid)
conductivities for $n=2$. 
using best-fit band parameters $t_1=0.25eV$, $t_2=0.15eV$
$t_3=0.03eV$ and $\Delta=0.3eV$ used in Fig. 1.
The $T=0$ conductivity was computed using an artificial
broadening of $0.1eV$ applied to the $B$ site.
The small contribution
to the $T=0$ conductivity arising from the minority
spin band is shown as the dash-dot line. }
\end{figure}

The conductivity corresponding to the parameters which maximize
$T_c$ (Fig 2) is shown in Fig. 6. We see that the different electronic
structure leads to a different optical conductivity:
the Drude absorbtion is weaker, and a peak corresponding
to excitation of carriers from $Fe$ to $Re/Mo$ is evident.

\begin{figure}
\includegraphics[width=3.0in]{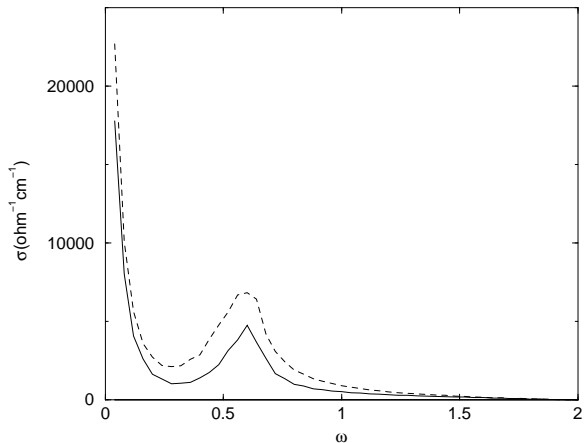}
\caption{$T=0$ (dashed) and $T>T_c$  (solid) optical 
conductivities for $n=1$ and band parameters
$t_1=0.25eV$, $t_2=-0.15eV$ and $\Delta=0$ corresponding
to the density of states shown in Fig 2.}
\end{figure}

The conductivity of  $Sr_2FeMoO_6$ has been 
measured by Jung and co-workers \cite{Jung02}. These
authors found a conductivity which was of roughly the Drude form,
(albeit with a rather larger scattering rate than we have used)
but additionally has a weak kink at a frequency of approximately
$0.6eV$. It is interesting to speculate that this kink is a signature
of the 'interband' feature which we found only for the 'antiphysical
parameters. A more detailed experimental investigation of the 
band structure may be warranted.

\section{Conclusions}

We have used the dynamical mean field method to determine the ferromagnetic
transition temperature, density of states and optical conductivity of a model
representing key physics (two dimensionality of band structure and strong
on-site interaction on $Fe$ site) of the double perovskite ferrimagnets
$Sr_{2}Fe(Mo/\operatorname{Re})O_{6}$. Our method can easily be generalized to
include the effects of mis-site disorder, or lattice distortions which
split the $t_{2g}$ levels. However, such generalization 
is not immediately warranted because
the calculated transition temperatures
are, at least for the parameters following from band structure calculations,
in qualitative disagreement with experimental data--in particular, the
calculation predicts that $Sr_{2}FeMoO_{6}$ is not ferrimagnetic,whereas
experiment indicates that it is with a $T_{c}$ is excess of $400K$, and
underpredicts the $T_{c}$ of $Sr_{2}Fe\operatorname{Re}O_{6}$ by a factor of
almost 4. 

The essential reason for this was found to be that the band theory
parameters imply that the mobile carriers reside mainly on the $non-Fe$ sites,
and hybridize weakly with these sites. Optical conductivity signatures of the
weak hybridization were demonstrated. The calculation indicates that
transition temperatures would be substantially raised if parameters are used
for which the added carriers are largely on the $Fe$ sites. An alternative
possibility is that an interaction omitted from the model is crucially
important; in particular that the magnetism should not be regarded
as arising from correlations on the $Fe$ site, but should
isntead be thought of more as a Stoner instability of the
band arising from the $Re/Mo$ states. The additional interaction
proposed in 
\cite{Ray01} would tend to produce this physics and an important 
next step would be to extend the methods developed here to the
treatment of this case.

\textit{Acknowledgements: }This work was supported by the University of
Maryland/Rutgers MRSEC (AJM and AC) and NSF-DMR-00081075 (KP) and
DARPA contract no. DAAD19-01-C-0060 (AC). We thank B. G.
Kotliar, S. Ogale, H.D. Drew, B. A. Jones and S. Cheong for helpful conversations.

\end{document}